\documentclass[conference]{IEEEtran}
\IEEEoverridecommandlockouts
\usepackage{cite}
\usepackage{amsmath,amssymb,amsfonts}
\usepackage{algorithmic}
\usepackage{graphicx}
\usepackage{textcomp}
\usepackage{xcolor}
\usepackage{hyperref}
\def\BibTeX{{\rm B\kern-.05em{\sc i\kern-.025em b}\kern-.08em
    T\kern-.1667em\lower.7ex\hbox{E}\kern-.125emX}}
    
\usepackage{mfirstuc}
\MFUnocap{are}
\MFUnocap{or}
\MFUnocap{etc}
\MFUnocap{of}
\MFUnocap{in}
\MFUnocap{the}
    
\usepackage{def}

\usepackage{tikz}
\usepackage{lipsum}

\newcommand\copyrighttext{%
  \footnotesize 978-1-6654-0615-4/21/\$31.00 \textcopyright 2021 IEEE}
\newcommand\copyrightnotice{%
\begin{tikzpicture}[remember picture,overlay]
\node[anchor=south,yshift=10pt] at (current page.south) {\parbox{\dimexpr\textwidth-\fboxsep-\fboxrule\relax}{\copyrighttext}};
\end{tikzpicture}%
}

\begin{document}
\bstctlcite{IEEEexample:BSTcontrol} 

\title{\capitalisewords{ Application of Gene Expression Programming in improving the event selection of the semi-leptonic top quark pair process\\}
\thanks{Latvian Council of Sciences project VPP-IZM-CERN-2020/1-0002.}
}

\author{\IEEEauthorblockN{ Andris Potrebko }
\IEEEauthorblockA{\textit{Centre of High-Energy Physics and Accele-} \\
\textit{rator Technologies, Riga Technical University}\\
Riga, Latvia \\
0000-0002-3776-8270}
\and
\IEEEauthorblockN{ Inese Polaka }
\IEEEauthorblockA{\textit{Department of Modelling and Simulation} \\
\textit{Riga Technical University}\\
Riga, Latvia \\
0000-0002-9892-7765}
}

\maketitle
\copyrightnotice

\begin{abstract}
Searches for \textit{Beyond the Standard Model} physics require probing the Standard Model with increased precision. One way this can be achieved is by improving the accuracy of the event selection classifiers. Recently, Gene Expression Programming (GEP) has been shown to provide complex yet easy to interpret classifiers in various fields. Previous attempts to apply GEP to high-energy physics (HEP), though limited by computational power available, achieved classifier accuracy of up to 95\%. In this paper, we demonstrate that a selection algorithm optimized by GEP and applied to the top-quark pair production process' semi-leptonic decay channel enables the increase of data purity for already highly pure data. Moreover, we explain how adding penalty cuts to the purity fitness function allows adjusting the optimized classifier to the needs of a specific measurement in terms of the size of the selected event sample and data purity. 
\end{abstract}


\section{Introduction}
Separation of the data of interest (signal) from unwanted noise (background) is a frequently encountered problem in physics studies performed by high-energy physics (HEP) experiments. 
Furthermore, expanding the searches for new physics and anomalies in the Standard Model require measurements of unprecedented accuracy, leading to the need for more accurate and faster data classifiers \cite{an2019,cerri2019, kekic2021}. Traditionally, data to background separation has been performed manually by implementing cuts on some data features (observables). More recently, modern multivariate classification methods and machine learning algorithms have been proven to provide better classification.
While methods like Support Vector Machines \cite{Vaiciulis2003}, Fisher Linear Discriminant \cite{Behnke2013} and, in most cases, Artificial Neural Networks \cite{Therhaag2013} are frequently used in HEP data selection, Evolutionary Algorithms (EA), including GEP, have had less attention\cite{link2005, Teodorescu2008}.

Unlike the closely related Genetic Algorithms (GA), GEP optimizes functions themselves, as opposed to numerical values or parameters of the model. This property allows the use of GEP in binary or multi-class classification problems, where it has been shown to provide reliable classifiers \cite{Zhou2003, Duan2006, Jedrzejowicz2019, Ferreira2006, Dehuri2008}. 
GEP has also been shown to be more effective than the related Evolutionary Programming (EP) \cite{Ferreira2006}. Finally, unlike Neural Networks, which are notorious for their black-box nature, classifiers provided by GEP are human-readable expression trees (ETs) and are useful in data mining. 
Previous results obtained using GEP in data selection of $K_S \rightarrow \pi^+ \pi^-$ events in $e^+e^-$ collisions provided classification accuracy of up to 95\% and were able to find valuable relations between input variables \cite{Teodorescu2008}. The results were, however, limited by the software and computation power available at that time.

In this paper, we review GEP as a method for improving event selection by applying it to the top/anti-top pair ($\ttbar$) production process in the semi-leptonic decay channel produced in proton-proton collisions in the LHC. We propose applying penalty cuts on data purity ($P$) to enable the use of $P$ as a fitness function. This new fitness function, $P_P$, allows the regulation of the trade-off between $P$ and the sample size of the selected data according to the specific experimental requirements. Moreover, we demonstrate that using $P_P$ to optimize a classifier on already pure data ($P>0.9$) provides more reliable results than the often used accuracy and sensitivity/specificity. The final classifier obtained was able to provide data with up to $P=0.96$, while preserving more than half of the initial data sample. 

The structure of the rest of this paper is as follows. In \secref{sec:GEP} we briefly review the fundamentals of the GEP algorithm. \secref{sec:Exp} introduces the fitness functions used in this study, the data set used and discusses the details of the implementation of the algorithm. The results of running the algorithm with each fitness function are shown in \secref{sec:res}.

\section{\label{sec:GEP} Gene Expression Programming}
GEP is an optimization technique that, similarly to other EAs, adapts the idea of natural evolution in evolving a population of the solutions of proposed problems (chromosomes, individuals) \cite{Ferreira2006}. The main differences between GEP and the closely related GA and EP are the chromosome structure, the mapping from a chromosome to a mathematical expression and the genetic operators used.

\begin{figure}[ht]
  \centering
  \includegraphics[width=0.95\linewidth]{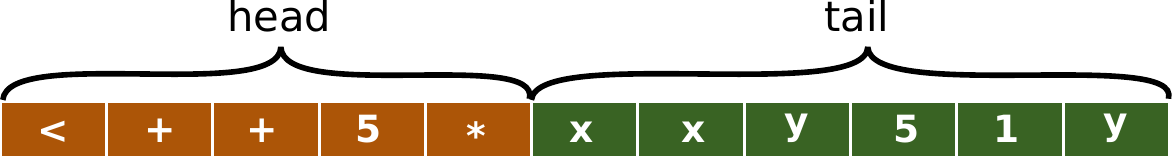}
  \caption{A schematic representation of an individual (chromosome) in GEP.}
  \label{fig:chromosome}
\end{figure}

The chromosome in GEP is divided into two parts. The first part (head) contains two kinds of primitives: terminals (constants or input variables) and functions. The second part (tail) contains only terminals. An example of a chromosome with a five-bit long head, containing functions $\{<,+,*\}$ and terminals $\{x,y,5,1\}$ is shown in \figref{fig:chromosome}.

The conversion of a chromosome into an ET starts with the first primitive and by applying the primitives following to the right as its arguments according to the number of arguments the function accepts (function arity). The process continues by filling up the next functions with their arguments until only terminals at the end of the branches remain. The purpose of the tail is to ensure that there are enough terminals to supply to each function.

All of the above applies to chromosomes consisting of one gene. It is also possible to build chromosomes consisting of several genes, which are linked by some user-defined linking function. 
\begin{figure}[ht]
  \centering
  \includegraphics[width=0.70\linewidth]{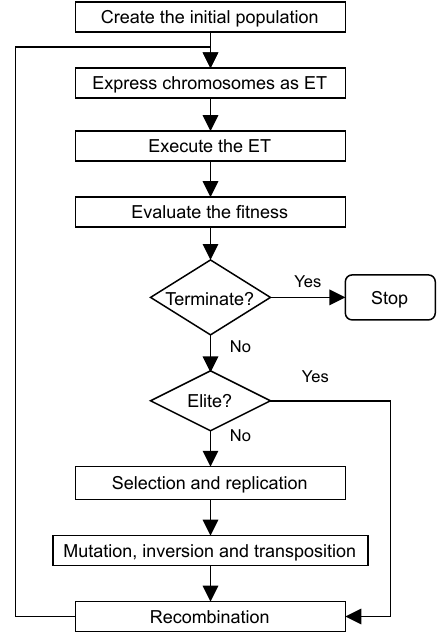}
  \caption{A flow chart for the GEP algorithm used.}
  \label{fig:GEP_flow}
\end{figure}

A flow chart showing the steps of the GEP algorithm used in this study is shown in \figref{fig:GEP_flow}. The termination criterion was chosen to be a predefined number of iterations.
As genetic operators, the classical set of GEP operators: roulette wheel selection, mutation, IS, RIS and gene inversion, one-point, two-point, and gene recombination, were chosen \cite{Ferreira2006}.
Numerical constants have been added using ephemeral numerical constants (ENC) method. The probabilities and rates for different GEP operators used in this study, as well as other settings, are discussed in \secref{sec:implementation}.




\section{\label{sec:Exp} Methodology }
\subsection{\label{sec:fitnessFun} Data selection, quality evaluation and fitness functions}

To represent the correctness of data classification in HEP, several quantities are used \cite{Teodorescu2008, Komiske2018, whiteson2009}. The most popular are:
\begin{itemize}
    \item classification accuracy ($Acc$),\\
    \begin{equation} \label{eq:accuracy}
        Acc = (TP+TN)/n,
    \end{equation}
    where $n$ is the total number of events, $TP$ $(N)$ represents the true positive (negative) rate and $FP$ $(N)$ represents the false positive (negative) rate; 
    \item signal efficiency ($Seff$),\\
    \begin{equation}
        Seff = TP/(TP+FN);
    \end{equation}
    \item background rejection ($Brej$),\\
    \begin{equation}
        Brej = TN/(TN+FP);
    \end{equation}
    \item and purity of the data sample ($P$),\\
    \begin{equation} \label{eq:purity}
        P = TP/(TP+FP).
    \end{equation}
\end{itemize}

For the last three of these, there exist trivial classification cases that produce the maximal possible score. For example, to get the maximal purity, the model can reject all the samples except one specific sample and thus get $FP=0$ and $P=1$. Since such trivial cases are usually more simple for an algorithm to find and hard to prevent, the last three quantities are not used as fitness functions for evolutionary algorithms.

To enable the use of purity as a fitness function we add a penalty for rejecting too many events or finding the trivial accept-all solution:
\begin{equation} \label{eq:P_Pen}
    P_P = \begin{cases}
          0 & \text{if }\; n_{acc}<n_{min} \;\text{ or\ }\; TN=0  \\
          P & \text{otherwise}\\
    \end{cases},
\end{equation}
where
\begin{equation}\label{eq:nacc}
    n_{acc}=\frac{TP+FP}{n}
\end{equation}
is the sample size of the accepted events relative to the initial sample size and $n_{min}$ is the minimal allowed sample size after the selection. In this way, $P_P$ offers a trade-off between what data purity is required for a particular experiment and how many events are needed for the analysis. It is also possible to define a more gradual cut by defining ($P_{ext}$) as a superposition
\begin{equation} \label{eq:purityPlysN}
    P_{ext} = \alpha\cdot P^{\beta} + (1-\alpha) \cdot n_{acc}^{\beta},
\end{equation}
where
$0\leq \alpha < 1$ and $\beta>0$ are constants.

It is common to combine $Seff$ and $Brej$, obtaining sensitivity/specificity ($SS$), as
\begin{equation} \label{eq:SS}
    SS = Seff \cdot Brej,
\end{equation} 
which enables the use of both of these quantities as fitness functions \cite{Ferreira2006}. Finally, when using $Acc$, it is possible to prevent the algorithm from getting stuck in trivial maxima of accepting or rejecting all events by applying a penalty
\begin{equation} \label{eq:AccPen}
    Acc_{P} = \begin{cases}
          0 & \text{if }\; TP=0 \; \text{ or }\; TN=0  \\
          Acc & \text{otherwise}\\
    \end{cases}
\end{equation}

The objective of this study is to test the usage of the fitness functions \eqref{eq:P_Pen}-\eqref{eq:purityPlysN} for GEP optimization of the event selection of data with already high purity. The performance is evaluated using the purity that the classifier can reach without decreasing $n_{acc}$ below 0.3.

The statistical error associated with a classification quantity $\varepsilon$ is calculated assuming that all $\varepsilon$ follow a binomial distribution:
\begin{equation} \label{eq:Error}
    \Delta \epsilon = z \sqrt{\frac{\varepsilon(1-\varepsilon)}{n}},
\end{equation}
where z is the $1-\frac{\alpha}{2}$ quantile of the Standard Normal distribution for the chosen confidence interval $1-\alpha$. For simplicity, we choose $1-\alpha=0.68$ ($1\sigma$ interval), where $z~=~1$. 

\subsection{The data set}
The data set used for this analysis was taken from the CERN Open Data portal --- a freely available access point for CERN data for research and education \cite{CMSOpenData2012}. The data set contains Monte Carlo (MC) samples of $\ttbar$ production, which serves as a signal in this study, and the relevant background. The subject of this classification is the frequently used semi-leptonic decay channel, i.e, $\ttbar$ events where one of the top quarks decays hadronically ($t\rightarrow W b \rightarrow q \overline{q} b$) and the other one leptonically ($t\rightarrow W b \rightarrow l \overline{\nu} b $)\cite{Chatrchyan2017}.

To filter out events in the semi-leptonic channel, a pre-selection of events with exactly one lepton with momentum $p_T>20$ GeV and pseudorapidity $\eta<2.4$, at least two-light jets with $p_T>30$ GeV and $\eta<2.4$ and at least two b jets with $p_T>30$ GeV and $\eta<2.4$ is applied. A higher-level feature --- the mass of the W boson candidate --- is built by calculating the combined mass of a two light jet combination that has the invariant mass closest to that of the W, $m_W=80$~GeV. In addition, seven low-level features are chosen, making eight variables in total, which are summarized in \tabref{tab:InputVar}.
More low-level features could have been chosen, but they were not considered helpful in distinguishing the semi-leptonic $\ttbar$ process and could lead to overtraining.

\begin{table}[ht!]
\centering
\caption{Set of variables used for GEP algorithm}
 \begin{tabular}{l|p{70mm} }
 Variable & Comment \\
 \hline
 $LepPt$  & Transverse momentum, $p_T$, of the lepton \\
 $LepEta$ & Rapidity, $\eta$, of the lepton \\
 $NLJets$   & Number of light jets \\
 $LJetPt0$  & $p_T$ of the most energetic jet from the W boson candidate  \\
 $LJetPt1$  & $p_T$ of the least energetic jet from the W boson candidate \\
 $LJetE0$  & Energy of the most energetic jet from the W boson candidate \\
 $LJetE1$   & Energy of the least energetic jet from the W boson candidate \\
 $Wmass$  & Mass of the W candidate\\
 \hline
\end{tabular}
\label{tab:InputVar}
\end{table}

The final dataset after the pre-selection consists of 1311 events that are shuffled and split into training and validation (evaluation) sets in an 80/20 proportion, or into 1036 and 275 events, respectively. Out of all the events, 1233 represent the signal (i.e, come from the 'MC:TTbar' data sample), while the remaining 78 events represent background. The initial purity of the training (validation) data sample is $P_0=0.940\pm 0.007 $ ($0.944 \pm 0.014$). This makes it a high-purity data sample and increasing this value is a challenging classification problem.


\subsection{Implementation and experimental parameters} \label{sec:implementation}

The classification is performed using Python 3.8 \cite{PythonSoftwareFoundation2}. Optimization is performed using a GEP framework \textit{geppy} \cite{geppy_2020, geppy_paper}, which is built on top of an evolutionary computation framework DEAP \cite{DEAP_JMLR2012}. The chosen input function set is a typical set used in GEP classification problems and consists of $+, -, \times, /, <, >, \leq, \geq, =, \neq$ \cite{Ferreira2006}.
Adding more complicated algebraic functions might help to find more complex relations between data input parameters. However, this adds a susceptibility to overtraining and has not previously shown large improvement for similar data\cite{Teodorescu2008}.

Geppy provides simplification options for the expressions obtained by GEP based on the sympy package \cite{meurer2017sympy}. However, these options were not defined for the case where the functions are both boolean and numeric. 
For this reason, a separate script, written in \textit{Mathematica}\cite{Mathematica} is used to simplify the output.

The default settings used in the study are shown in \tabref{tab:settings}. The rates for the genetic operators are kept as the recommended ones as they have been shown to work with a wide range of classification problems \cite{Ferreira2006, Teodorescu2008}.
\begin{table}[ht]
\caption{The default settings used in the computation
}
\centering
 \begin{tabular}{l|c| }
 Number of runs & 20 \\
 Number of generation & 500 \\
 Population size & 40 \\
 Head length & 7  \\
 Number of genes & 2 \\
 Linking function & AND \\
 Input variables & \tabref{tab:InputVar}\\
 Number of elites & 1\\
\end{tabular}
 \begin{tabular}{|p{28mm}|c| }
 Inversion rate & 0.1 \\
 Mutation rate & 0.05 \\
 IS transposition rate & 0.1 \\
 RIS transposition rate & 0.1  \\
 One-point recomb. rate & 0.4 \\
 Two-point recomb. rate & 0.2 \\
 Gene crossover rate & 0.1  \\
 ENC mutation rate & 0.1 \\
 \end{tabular}
\label{tab:settings}
\end{table}

\section{Results} \label{sec:res}
Four fitness functions, \eqref{eq:P_Pen}-\eqref{eq:purityPlysN}, were tested for GEP optimization on the given data sample. Settings shown in \tabref{tab:settings} were used for all cases.
The performance of each fitness function was compared in terms of the achieved $P$ on the validation data for the fittest individual.

\subsection{Performance of sensitivity/specificity fitness function}\label{sec:SS}

\begin{figure}[ht]
  \centering
  \includegraphics[width=0.9\linewidth]{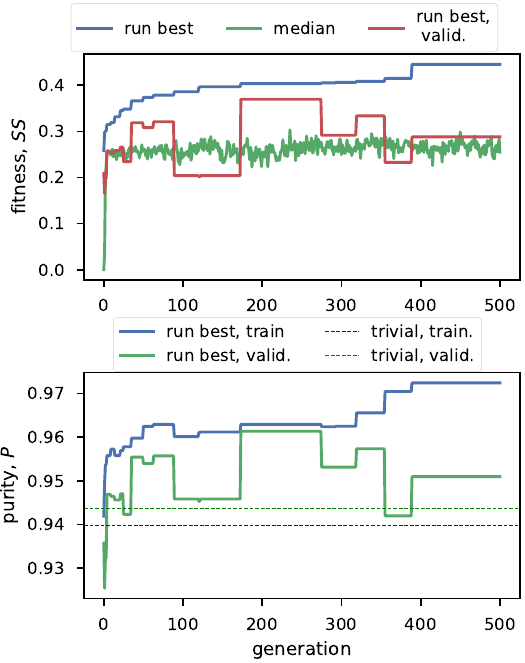}
  \caption{Change of SS fitness (upper) and purity (lower) values of the fittest individual in all the population and all runs with the number of generations.}
\label{fig:exampleStats_SS}
\end{figure}

First, optimization was performed with the $SS$ fitness function, \eqref{eq:SS}.
The algorithm was repeated for 20 runs with different starting populations.
The convergence of fitness and $P$ for the fittest individual evaluated on both training and validation sets in all the runs is shown in \figref{fig:exampleStats_SS}.
The median fitness value of the population averaged over all the runs is also depicted.
A clear correlation is visible between the $SS$ fitness function and $P$ for both the training and validation samples.
This shows that $SS$ is a good candidate for a fitness function given increased data purity as the objective.

While both overfitting and statistical fluctuation due to the small data sample are present, purity of the validation sample increases over the trivial (except-all) purity, $P_0$.
Due to the statistical fluctuation of the fitness on the validation sample, it is not clear how to choose a generation at which to terminate the algorithm to prevent overfitting. For this reason, the performance was measured as $P$ on the final individual, i.e, the fittest individual on the training set.
Purity values reached were $P=0.972\pm 0.005$ and $P=0.951\pm 0.015$ for the training sample and the validation sample, respectively.
While within the error intervals $P$ for the validation sample still includes $P_0 = 0.944\pm0.014$, it provides an improvement for the central value.
Relative sample size for the training sample was $n_{acc}=0.589\pm 0.015$, providing an experimentally suitable data size.

\subsection{Performance of accuracy fitness function}\label{sec:Acc}

In this section, the runs of the previous section were repeated using the $Acc_P$ fitness function, \eqref{eq:AccPen}. The purity for the fittest individuals reached $P=0.941\pm 0.007$ and $P=0.944\pm 0.015$ for the training sample and the validation sample, respectively, and the sample size for the training sample was $n_{acc}=0.999$.
The results show that, while the penalty in \eqref{eq:AccPen} prevented the algorithm from getting trapped in a trivial extreme, the algorithm now gets stuck \textit{around} the trivial extreme.
In other words, the best solution found is one where almost all the samples get accepted except one.

These results can be seen as contradicting Ref.\cite{Teodorescu2008}, where $Acc$ showed non-trivial, high-acceptance results for data with a wide range signal/background ratios from 0.25 to 5.
It was found that $Acc$ even provided a slight improvement in accuracy when using data with a larger signal/background ratio, that is, when using data with a signal/background ratio of 5 instead of 1.
Nevertheless, as seen in \secref{sec:SS}, increasing purity by a few percentiles required discarding almost $50\%$ of the data.
Since $Acc$ attempts to remove $FP$ and $FN$ samples with equal weights, it might fail on hardly classifiable data where purity improvement might result in discarding many positive samples and in this way increasing the number of $FN$ samples.

\subsection{Performance of purity with a cut on number of samples}\label{sec:P_P}
\begin{figure}[ht]
  \centering
  \includegraphics[width=0.9\linewidth]{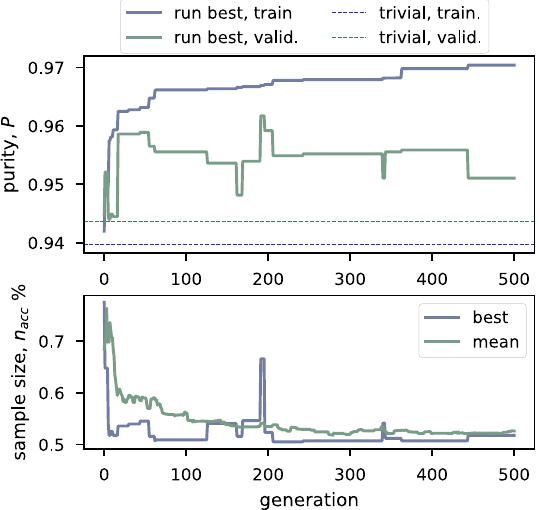}
  \caption{Change of purity (upper) and relative samples size, $n_{acc}$, (lower) values of the fittest individual in all the runs with the number of generations. For $n_{acc}$ also the mean over the runs is shown. The fitness function used is $P_P$ (see, \eqref{eq:P_Pen}) with $n_{min}=0.5$.}
  \label{fig:exampleStats_P}
\end{figure}

Calculations were repeated with the $P_P$ (see, \eqref{eq:P_Pen}) fitness function with different values of $0.1<n_{min}<0.9$. \figref{fig:exampleStats_P} shows the change of $P$ and $n_{acc}$ with respect to the generation number for an example case with $n_{min}=0.5$. For $n_{acc}$ the value for the fittest individual in all the runs, as well as the mean of the fittest individuals in all the runs are shown. It is visible that, as the population fittest individuals evolve, $n_{acc}$ converges to $n_{min}$. Similarly, the individual providing the highest $P$ is mostly the one that rejects data the most. 
This is a manifestation of the fact that reducing $FN$ samples also causes a reduction in the $TP$ sample, leading to a  smaller $n_{acc}$.

\begin{table}[ht]
\caption{Purity and sample size of the fittest individual over all the runs for different $n_{min}$ values  }
\centering
 \begin{tabular}{l|c|c|c| }
 $n_{min}$ & final $P$, train. set & $n_{acc}$, training set & final $P$, valid. set\\
 \hline
 0.1 & $1.000 \pm 0.000$  & $0.107$  & $0.956\pm 0.013$ \\
 0.3 & $0.979\pm0.004$ & $0.313 $ & $0.947 \pm 0.014$ \\
 0.5 & $0.970\pm 0.005$ & $ 0.518  $ & $0.951 \pm 0.013$ \\
 0.7 & $0.968 \pm 0.005 $ & $0.713$ & $0.952 \pm 0.013 $ \\
 0.9 & $0.950 \pm 0.007 $ & $0.938$ & $0.943 \pm 0.014$ \\
\end{tabular}
\label{tab:Res_P}
\end{table}
The trade-off between increasing $P$ and preserving the number of samples is also visible in \tabref{tab:Res_P}, where $P$ and $n_{acc}$ of the fittest individuals are shown for calculations with different $n_{min}$.
In all the cases, $n_{acc}$ approaches $n_{min}$ as the individuals benefit from the opportunity of making increasingly larger cuts.
The fact that $P$ for the validation and training samples do not obviously correlate is a consequence of overfitting and still has to be properly dealt with. 

\subsection{Performance of purity with a gradual cut on sample size} \label{sec:P_ext}
\begin{figure}[ht]
\centering
  \includegraphics[width=0.95\linewidth]{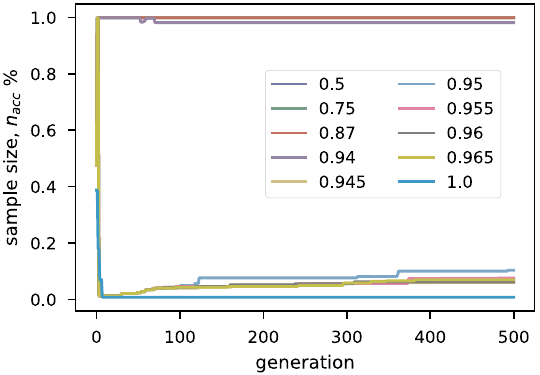}
  \caption{Change of $n_{acc}$ with the number of generations using $P_{ext}$ as a fitness function with $\beta=1$ for different $\alpha$ values.}
  \label{fig:P_ext}
\end{figure}
Finally, the calculation was repeated with $P_{ext}$, (see, \eqref{fig:P_ext}) as a fitness function. The parameter $\beta$ was chosen as $\beta=1$ and $\alpha$ varied within the interval $0.5\leq\alpha\leq1$. The evolution of $n_{acc}$ (see \figref{fig:P_ext}) shows that already after a few generations almost all of the attempts got trapped in maxima (in case of $\alpha=1$, the global maximum) of two kinds depending on if either $P$ part or $n_{acc}$ part dominates in \eqref{fig:P_ext}. Samples with large $\alpha$ ($\alpha > 0.945$), where the $P$ part dominates, got stuck in accepting only a few samples, while samples with small $\alpha$ got stuck in accepting almost all the samples. While the slight improvement in the trials with $0.950\leq\alpha \leq 0.965$ shows that there might be a good balance point between the two effects, it does not seem to be easily found and should depend on the initial value of $P$ and the number of samples $n$.

\subsection{Structure of optimized individuals}
Some examples of the optimized classifiers are shown in \tabref{tab:ResFunctionSets2}.
The expressions contain relations, like $-22.17\cdot(-43.26 + LJetE1) \geq LJetPt1^{2}$ resembling a requirement on the invariant mass of the second light jet, or cuts on particle momenta and energies like, $LepPt < LJetPt0$, resembling selection on hard jets.
While the specific physics that GEP is pointing to is not clear, the selection is similar to how it could have been done manually and involves variables where higher-level relations could be found.
GEP could be supported by providing it with already known physical relations and cuts on higher-level physical features, such as the $\chi^2$ value for the fit of the $t$ quark candidate.
\begin{table}[ht]
\caption{The fittest individuals of GEP optimization for different fitness functions. The numerical values are rounded up to four significant digits.}
\centering
 \begin{tabular}{p{1.5cm}|p{6.4cm} }
 Fitness function & The most fit individual simplified \\
 \hline
 $SS$ & $(-22.17\cdot(-43.26 + LJetE1) \geq LJetPt1^2)$
 $\text{AND}\; ((LepPt < LJetPt0) \geq -LepEta/NLJets)$ \\
 $P_P$, $n_{min}=0.5$ & $(LepPt + 39.60\cdot LJetE0 + LJetPt0^2 \geq LJetE1\cdot $
 $LJetPt0) \;\text{AND}\; (LepEta + LJetE0 + (1167 +$
 $LepPt)\cdot Wmass \geq 1167\cdot LJetPt0)$ \\
  $P_P$, $n_{min}=0.9$ & $(2\cdot LepPt + Wmass + (LJetE0 \geq$
  $LJetE1) \geq LJetE0) \;\text{AND}\; (LJetPt0 + Wmass \geq 24.34 +$
  $LJetPt1 + NLJets)$ \\
\end{tabular}
\label{tab:ResFunctionSets2}
\end{table}

Adding additional variables would increase the vulnerability to overfitting.
The irrelevant or correlated variables could be removed by, for instance, Bayesian hyperparameter optimization \cite{Klein2015}.

\section{Conclusion and Outlook}

In this initial study, the performance of various GEP fitness functions was tested in optimizing the data selection algorithm for high-purity HEP data.
A quantitative evaluation of the performance was complicated due to the small data sample used leading to large statistical errors and overfitting.
While none of the calculations could increase $P$ above the error intervals of $P_0$, nearly all calculations showed an increase in terms of a central value (see, e.g, \figref{fig:exampleStats_SS} and all cases with $n_{min}>0.1$ in \tabref{tab:Res_P}).
Still, this result is left to be verified in further studies with increased sample size.

As shown in \secref{sec:P_P}, while $P$ itself is not usable as a fitness function in GEP, the use of it can be enabled by applying appropriate penalties on choosing $n_{acc}<n_{min}$.
While both fitness functions $P_P$ and $SS$ provided similar results, $P_P$ can be more desirable for two reasons.
Firstly, for experiments where $P$ is the main measure of the quality of data selection, using $P_P$ directly does not require evaluating the correlation of $P$ and the fitness function used on the given data.
Secondly, $P_P$ allows to tune $P$ and $n_{acc}$ of the selected data sample.
While similar cuts for $n_{acc}<n_{min}$ can be applied also on $SS$, there is no guarantee that $n_{acc}$ would converge to $n_{min}$. 

In practical cases, the process of the data selection optimization could take place by first choosing the necessary statistics $n_{min}$ that, assuming a possible final $P$, would provide the desired statistical uncertainty.
If the predicted $P$ is not reached, optimization could plausibly be repeated with a different $n_{min}$ until the desired values of both $n_{min}$ and $P$ are reached.

It is possible to add different, not-abrupt penalties on $P$. Nevertheless, the attempts of using $P_{ext}$ in \secref{sec:P_ext} yielded no improvement, and possible penalty mechanisms should still be researched both theoretically and experimentally.

As suggested by the results of \secref{sec:Acc}, the performance of GEP can vary considerably on the data studied and its initial purity. This shows the need to check the performance of different fitness functions on a wider range of signal/background ratio data as initially done in \cite{Teodorescu2008}. Similarly, early stopping or other qualitative techniques to prevent overfitting should be applied when extending this study \cite{Tuite2011}.

\section*{Acknowledgments}
We would like to thank the CERN Open Data portal for providing the open-access data which enabled this study.
We also thank the Latvian Council of Sciences for the provision of funding for the State Research Programme project VPP-IZM-CERN-2020/1-0002, under which this study was performed.
Finally, our thanks go to the Riga Technical University Faculty of Computer Science and Information Technology.

\bibliographystyle{IEEEtran}
\bibliography{bib/Studies-EvolutionaryAlgorithms_url, bib/BibByHand, bib/Studies-GA}

\end{document}